\documentclass[prc,twocolumn,showpacs,preprintnumbers,amsmath,amssymb]{revtex4}

\usepackage{graphicx}% Include figure files
\usepackage{dcolumn}% Align table columns on decimal point
\usepackage{bm}% bold math
\input{epsf.tex}

\def\be{\begin{equation}}
\def\ee{\end{equation}}
\def\ba{\begin{array}}
\def\ea{\end{array}}
\def\bea{\begin{eqnarray}}
\def\eea{\end{eqnarray}}
\begin{document}

\title{Quest of $^{37}Mg$ halo structure using Glauber model and microscopic relativistic mean field densities}

\author{\large Mahesh K. Sharma}\thanks{{maheshphy82@gmail.com}}
\affiliation{School of Physics and Materials Science, Thapar University, Patiala - 147 004, Punjab, India}
\author{\large R. N. Panda}
\affiliation{Department of Physics, ITER, Siksha O Anusandhan University, Bhubaneswar-751 030, India}
\author{\large Manoj K. Sharma}
\affiliation{School of Physics and Materials Science, Thapar
University, Patiala - 147 004, Punjab, India}
\author{\large S. K. Patra}
\affiliation{Institute of Physics, Sahcivalaya marg Bhubaneswar-751 005, India}

\date{\today}% It is always \today, today,

\begin{abstract}
We have studied the ground state properties (binding energy and charge radius) using relativistic
mean filed formalism (RMF) for Mg-isotopes in the valley of stability to drip line region.
The RMF densities have been analyzed in context of reaction dynamics.
The calculated results of $^{24-40}$Mg+$^{12}$C reactions at projectile energy 240 AMeV using Glauber
model with the conjunction of densities from relativistic mean field formalism are compared with experimental data.
We found remarkable agreement of estimated values of reaction cross sections with experimental data except for $^{37}$Mg isotope.
In view of this, the halo status of $^{37}$Mg is examined through higher magnitude of rms
radius and small value of longitudinal momentum distribution. Finally, an effort is made to explore the structure
of $^{37}$Mg halo candidate using Glauber few body formalism.
\end{abstract}

%Use showkeys class option if keyword

\maketitle

\section{\label{sec:level1} INTRODUCTION}
Since last few decades, the advent in radioactive ion beam facilities, the drip line physics and
structure of drip line nuclei got large attention by the nuclear science community.
The advancement in these facilities enrich our knowledge toward the new exotic phenomena such as
one and two neutron/proton halo, bubble effect in densities, vanishing of shell closer effect
at the drip line region and identification of some new magic numbers for the deformed configuration etc.
The understanding of above mention concepts strengthen our understanding in the island of inversion.
One of the most exotic phenomenon that have been exploited extensively toward drip line is halo status of nuclei.
The halo structure originates due to extremely weak bound nucleons which decouple from the nuclear core.
The interaction cross sections of such nuclei like $^{6,8}He$,
$^{11}Li$ and $^{11,14}Be$ \cite{tani92} are anomalously high due to large magnitude of consolidated
root mean square (rms) radii.
The one neutron/proton breakup of halo projectiles with stable target at
near/far barrier energies is one of the tool to investigate the structure of such systems \cite{paes2012}.
The measurement of longitudinal momentum distribution for $^{16-18}$C isotopes after one-neutron
breakup from $^{17-19}$C isotopes using the fragment separator observed narrow value of full width half
maximum (FWHM) for $^{18}$C ($\sim$ $44.3\pm5.9$ MeV/c),
which indicates that $^{19}$C is a one-neutron halo\cite{Bazin1995}. In addition to this, the measurement of nuclear
reaction cross-sections for
$^{19,20,22}C$ \cite{tana10,Koba2011} show that the drip-line nucleus
$^{22}C$ has a halo structure. The one and two neutron removal cross sections
and momentum distribution give indication for the halo status of a nuclear system.
It is relevant to mention that $^{22}C$ has N = 16 which is a new magic number in neutron-rich
region \cite{ozawa2000,Tani2001} and forms a Borromean halo structure
($^{21}C$ is unstable). It is of large interest to study the existence of bound states of two nucleons
(neutrons or protons) with core in the Borromean structure, which have not appeared in vacuum.
Soon after $^{31}Ne$ was included in the family of neutron
halo. The measurement of interaction cross sections for Ne
isotopes from
the stability line to neutron drip
line at 240 MeV/nucleon energy \cite{Hama2010}, show the dependence of mass number for
$^{27-32}$Ne
isotopes, which have been explained by nuclear deformations. The enhancements of interaction cross sections  particularly
for $^{29}$Ne and $^{31}$Ne have been quoted by an
s-dominant halo structure of $^{29}$Ne and s- or p-orbital halo in $^{31}$Ne
\cite{Takechi2012}.
The isotope $^{31}Ne$ having neutron number N=21, seem to break
the shell closer structure. As a consequence, a large value of deformation
associated with the strong intruder configuration put in at "island of
inversion".\\
Recently, M. Takechi et al. \cite{takechi2015}, have measured the data of reaction cross
section for Mg isotopes at energy 240 A MeV at RIBF and RIKEN. In this study, they have observed large
reaction cross section value for $^{37}$Mg as compared to other isotopic chain and consequently
predicted $^{37}$Mg to have a halo structure. In another study S. Watanabe \cite{Watanabe2014} concluded from their
fully microscopic double folding frame work with the AMD densities, that $^{37}$Mg exhibits deformed halo structure.
In our earlier work we explored the structure of some isotopes of Ne, Mg and
Si \cite{panda2014}. The identification of this new candidate  as halo encourage us to review our study for Mg
isotopes. Therefore we have made a motivated effort to study the reaction dynamics of Mg isotopes
at energy 240 AMeV. A sincere effort is made to analyze the structural features of
$^{37}$Mg. The reaction/interaction cross section, angular elastic differential cross section,
one nucleon removal cross section and angular momentum distributions parameters
are exercised for the analysis of such nuclear systems. \\
The section II contains a brief description of
Glauber formalism. Section III describe the calculations and results,
and finally a summary and conclusions is described in the section IV.

\section{The Formalism}

We use the well known Glauber approach to investigate
reaction dynamics \cite{glau55,Karo75,Chauvin1983}. The study of reaction
dynamics in the framework of this approach strongly depends on the densities of the projectile and target nuclei,
and we have used the microscopic relativistic mean field (RMF) densities with
NL3 parameter \cite{lala97}.
\subsection{Glauber Model}
\subsubsection{Reaction cross section}
The theoretical formalism to study the reaction cross sections
using the Glauber approach has been given by R. J. Glauber \cite{glau55}.
The standard Glauber form for total reaction cross
sections is expressed as \cite{glau55,Karo75}
\begin{equation}
\sigma_R=2\pi\int_0^\infty b[1-T(b)]db,
\end{equation}
where 'T(b)' is the Transparency function with impact parameter
'b'. The function T(b) is calculated by
\begin{equation}
T(b)=exp[-\sum_{i,j}\sigma_{ij}\int\overline{\rho_{tj}}(s)
\overline{\rho_{pi}}(|\overrightarrow{b}-\overrightarrow{s}|)
\overrightarrow{ds}].
\end{equation}
Here, the summation indices i, j run over proton and neutron and
subscript 'p' and 't' refers to projectile and target
respectively. $\sigma_{ij}$ is the experimental nucleon-nucleon
reaction cross-section which depends on the energy. The z-
integrated densities are defined as
\begin{equation}
\overline{\rho}(\omega) =\int_{-\infty}
^\infty\rho(\sqrt{w^{2}+z^{2}}) dz,
\end{equation}
with $\omega^2=x^2+y^2$.
Initially Glauber model was designed for the
high energy approximation. However, it was found to work reasonably
well for both the nucleus-nucleus reaction and the differential
elastic cross-sections
over a broad energy range \cite{Chauvin1983,Buenerd1984}.
The modified transparency function T(b) is given by
\begin{equation}
 T(b)=exp[-\int_p\int_t\sum_{i,j}[\Gamma_{ij}(\overrightarrow{b}-\overrightarrow{s}+\overrightarrow{t})]
 \overrightarrow{\rho_{pi}}(\overrightarrow{t})
\overrightarrow{\rho_{tj}}(\overrightarrow{s})
\overrightarrow{ds}\overrightarrow{dt}].
\end{equation}
The profile function $\Gamma_{NN}$ for optical limit approximation is
defined as
\begin{equation}
\Gamma_{NN} =\Gamma_{ij}(b_{eff}) =\frac{1-\iota\alpha_{NN}}{2\pi
\beta^2_{NN}}\sigma_{NN} exp(-\frac{b^2_{eff}}{2\beta^2_{NN}}),
\end{equation}
for finite range and
\begin{equation}
\Gamma_{NN} =\Gamma_{ij}(b_{eff}) =\frac{1-\iota\alpha_{NN}}{2}\sigma_{NN} \delta (b),
\end{equation}
for zero range
with
$b_{eff}$=$|\overrightarrow{b}-\overrightarrow{s}+\overrightarrow{t}|$, $\overrightarrow{b}$ is the impact parameter. Where
$\overrightarrow{s}$ and $\overrightarrow{t}$ are the dummy variables for integration
over the z-integrated target and projectile densities. The
parameters $\sigma_{NN},\alpha_{NN}$ and $ \beta_{NN}$ usually
depend upon the proton-proton, neutron-neutron and proton-neutron
interactions. Here $\sigma_{NN}$ is the total nuclear reaction cross section of NN
collision, $\alpha_{NN}$ is the ratio of the real to the imaginary part
of the forward nucleon-nucleon scattering amplitude and $\beta_{NN}$ is the slope
parameter. The slope parameter determines the fall of the angular
distribution of the N-N elastic scattering.
\subsubsection{Angular elastic differential cross section}
The nucleus-nucleus elastic scattering amplitude is
written as
\begin{equation}
F(q)=\frac{\iota K}{2\pi} \int db e^{\iota q.b}(1-e^{\iota \chi(b)}).
\end{equation}
At low energy,  this model is modified in order to take care of finite
range effects in the profile function and
Coulomb modified trajectories. The elastic scattering amplitude including
the Coulomb interaction is expressed as
\begin{equation}
F(q)=e^{\iota \chi_{s}}\{F_{coul}(q)+\frac{\iota K}{2\pi}
\int db e^{\iota q.b+2\iota \eta \ln(Kb)}(1-e^{\iota \chi(b)})\},
\end{equation}
with the Coulomb elastic scattering amplitude
\begin{equation}
F_{coul}(q)=\frac{-2 \eta K}{q^2}exp\{-2 \iota \eta \ln(\frac{q}{2K})
+2\iota arg \Gamma(1+\iota \eta)\},
\end{equation}
where $K$ is the momentum of projectile and $q$ is the momentum transferred
from the projectile to the target. Here $\eta=Z_P Z_T e^2/\hbar v$ is the
Sommerfeld parameter, $v$ is the incident velocity of the projectile,
and $\chi_s=-2\eta \ln(2Ka)$ with $a$ being a screening radius.
The elastic differential cross section is given by
\begin{equation}
\frac{d\sigma}{d\Omega}=|F(q)|^2.\\
\end{equation}
\begin{equation}
\frac{d\sigma}{d\sigma_R}=\frac{|F(q)|^2}{|F_{coul}(q)|^2}.
\end{equation}

\subsubsection{one nucleon removal cross section}
The one nucleon removal cross section, $\sigma_{-N}$, may be defined as
\begin{equation}
\sigma_{-N}=\Sigma_{c}\int dk \sigma_{a=(k,g=0),c}
\end{equation}
we assume that the core remains in its ground state, g=0.
The one nucleon removal cross section consists of both elastic and
inelastic part and can be calculated by
\begin{equation}
\sigma_{-N}=\sigma^{el}_{-N}+\sigma^{inel}_{-N}.
\end{equation}
The cross section due to the elastic breakup process is given by
\begin{eqnarray}
\sigma^{el}_{-N}&=&\int db \{\langle\phi_0|e^{-2Im\chi_{CT}(b_c)-2Im\chi_{NT}(b_c+s)}|\phi_0\rangle\nonumber\\
&&-|\langle\phi_0|e^{\iota\chi_{CT}(b_c)+\iota\chi_{NT}(b_c+s)}|\phi_0\rangle|^2\}.
\end{eqnarray}
While the cross section from inelastic breakup is given by
\begin{eqnarray}
\sigma^{inel}_{-N}&=&\int db\{\langle\phi_0|e^{-2Im\chi_{CT}(b_c)}\nonumber\\
&&-e^{-2Im\chi_{CT}(b_c)-2Im\chi_{NT}(b_c+s)}|\phi_0\rangle|\}.
\end{eqnarray}
\subsubsection{Longitudinal momentum distribution}
The momentum distribution of core after the inelastic breakup of Projectile reads as:
\begin{eqnarray}
\frac{d\sigma^{inel}_{-N}}{dP}=\int \frac{dq}{K^2} \sum_{c\neq0}\int dk \delta (P-\frac{A_C}{A_P} \hbar q+\hbar K )|F_{(K,0)_c} (q)|^2 .
\end{eqnarray}
The Scattering wave function of nucleon is approximated by a plane wave reduced
to
\begin{eqnarray}
\frac{d\sigma^{inel}_{-N}}{dP_{||}}=\int db_N (1-e^{-Im \chi_{NT}(b_N)})\times\frac{1}{(2\pi\hbar)^3} \frac{1}{2j+1} \nonumber\\
\sum_{m m_s}|\int dr e^{\frac{i}{h}P.r}\chi_{\frac{1}{2}m_s} e^{i\chi{CT}(b_N-S)\varphi_{nljm}(r)}|^2,
\end{eqnarray}
 where $b_N$ stands for the impact parameter of valence nucleon with respect to the target. The longitudinal momentum distribution
 obtained by the integration of above equation over  transverse component of momentum $P_{\bot}$, gives.
\begin{eqnarray}
\frac{d\sigma^{inel}_{-N}}{dP_{||}}=\int dP_{\bot}\frac{d\sigma^{inel}_{-N}}{dP}\nonumber
\end{eqnarray}
\begin{eqnarray}
&=&\frac{1}{2\pi\hbar}\int db_N(1-e^{-2I_m \chi_{NT}(b_N)})\int ds e^{-2I_m \chi_{CT}(b_{N}-s)}\nonumber\\
&& \times \frac{1}{(2l+1)}|\int dz e^{\frac{\iota}{\hbar}P_{||}z} u_{nlj}(r) Y_{l m_l}(\hat{r})|^2.
\end{eqnarray}

\subsection{Relativistic mean field formalism}
The relativistic mean field approach is well documented in Refs. \cite{pannert1987,baguta1977,ring1996,patra1991,Del2001}.
The basic ingredient of RMF model is the relativistic Lagrangian density for a nucleon-meson many body
system which is defined as \cite{patra1991}
\begin{eqnarray}
{\cal L}
&=& \overline{\psi}_i(i\gamma^{\mu}\partial_{\mu}-M)\psi_i+{\frac{1}{2}}\partial^{\mu}\sigma\partial_{\mu}\sigma\nonumber\\
&&-\frac{1}{2}m^{2}_{\sigma}\sigma^{2}-\frac{1}{3}g_2\sigma^3-\frac{1}{4}g_3\sigma^4-g_s\overline{\psi}_i\psi_i\sigma \nonumber\\
&&-{\frac{1}{4}}\Omega^{\mu\nu}\Omega_{\mu\nu}+{1\over{2}}m_{w}^{2}V^{\mu}V_{\mu} \nonumber\\
&&-g_\omega\overline{\psi}_i\gamma^{\mu}\psi_iV_{\mu}-{\frac{1}{4}}\vec{B}^{\mu\nu}.\vec{B}_{\mu\nu} \nonumber\\
&&+{1\over{2}}m_{\rho}^{2}{\vec R^{\mu}}.{\vec{R}_{\mu}}-g_\rho\overline{\psi}_i\gamma^{\mu}\overrightarrow{\tau}\psi_i.\overrightarrow{R^{\mu}} \nonumber\\
&&-{1\over{4}}F^{\mu\nu}F_{\mu\nu}-e\overline{\psi}_i\gamma^{\mu}\frac{(1-\tau_{3i})}{2}\psi_iA_{\mu}.
\end{eqnarray}
Here $\sigma$, $V_{\mu}$ and $\overrightarrow{R}_{\mu}$ are the fields for $\sigma$-,
 $\omega$- and $\rho$-meson respectively. $A^{\mu}$ is the electromagnetic
field. The $\psi_i$ are the Dirac spinors for the nucleons whose third
component of isospin is denoted by $\tau_{3i}$. $g_s$, $g_\omega$,
$g_\rho$ and $\frac{e^2}{4\pi}=\frac{1}{137}$ are the coupling constants
for the linear term of $\sigma$-, $\omega$- and $\rho$-mesons and photons
respectively. $g_2$ and $g_3$ are the parameters for the non-linear
terms of the $\sigma$-meson.
M, $m_\sigma$, $m_\omega$ and $m_\rho$ are the masses of the nucleons,
$\sigma$-, $\omega$- and $\rho$-mesons, respectively. $\omega^{\mu\nu}$,
$\overrightarrow{B}^{\mu\nu}$ and $F^{\mu\nu}$ are the field tensors for
the $V^{\mu}$, $\overrightarrow{R}^{\mu}$ and the photon fields, respectively.
The quadrupole moment deformation parameter $\beta_2$, root mean square radii
and binding energy are evaluated using the standard relations \cite{pannert1987}.
The nuclear density $\rho={\sum}^{A}_{i=1}\psi^{\dag}_i\psi_i$ is obtained by
solving the equation of motion obtained from the above Lagrangian
%\cite{pannert1987,patra1991}.
The values of the parameters for NL3 are given as \cite{lala97}
$g_{s}$=10.217, $g_{\omega}$=12.868,
$g_{\rho}$=4.574, $g_{2}$ = -10.431 ($fm^{-1}$), $g_3$ = -28.885, and M=939,
$m_{\sigma}$=508.194, $m_{\omega}$=782.501, $m_{\rho}$=763.0 in MeV.

\section{CALCULATIONS AND DISCUSSION}
The ground state properties of Mg-isotopes are calculated using microscopic relativistic mean field formalism
Fig 1 shows the calculated values of binding energy (B. E.) and charge radius ($r_c$) of considered
set of isotopes of Mg. The lower panel of the figure shows
the calculated values of
B.E, which find nice comparison with experimental data which looks
nice agreement with each others. A deep inspection of the figure shows that the B. E.'s slightly under estimate up to the $^{29}$Mg and overestimate beyond
$^{32}$Mg isotopes.
While the upper panel of figure compare the calculated value of rms charge radius ($r_c$) of considered
set of isotopes with the experimental data. The calculated values of $r_c$ are underestimated in comparison to experimental data.
The success of relativistic mean field formalism depends on the appropriate choice of densities and we
have used RMF(NL3) densities as an input of Glauber model.
\begin{figure}
%\vspace{-1.7cm}
%\hspace{-0.7cm}
\includegraphics[width=0.8\columnwidth,clip=true]{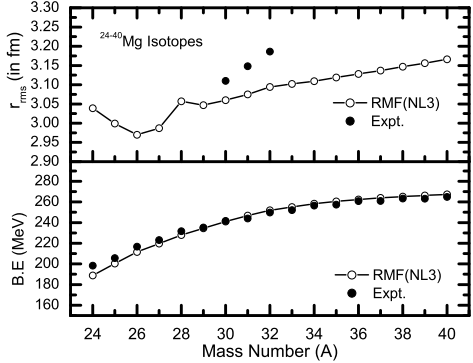}
\label{Fig.1}\vspace{-0.7cm} \caption{The values of B.E in MeV and Charge radius of Mg isotopes after obtained from RMF(NL3) as a function of mass number (A). The experimental data are also given for comparison whichever be available.}
\end{figure}

In the measurement of reaction parameter through Glauber formalism, one of the input for evaluation of
profile function in Glauber model is
its energy as well as isospin dependent parameters. The values of these parameter
at $E_{Proj}$= 240 AMeV are $\sigma_{NN}$= 3.266868 $(fm^2)$, $\alpha_{NN}$=0.6800303
and $\beta_{NN}$= 0.097843707 $(fm^2)$. These values have been estimated by spline
interpolation from Ref. \cite{horiu07}.
Other important inputs of Glauber model are the densities of the
projectile and the target nuclei. Fig. 2 shows the relativistic mean filed (RMF) densities of projectiles
with NL3 parameter set, which shows the density profile as a function of radial distance for $^{24-40}$Mg nuclei.
Here the nucleonic densities distribution are of larger values at the center and goes on decreasing as the radius increases.
where the small depletion in densities also appear at the center for these nuclei.
One may also observed from the figure that the skin effect increases with increase of isotopic mass number. \\
These densities can be feed as an input of Glauber model after converting
into spherical equivalent of it in terms of Gaussian coefficients. We have converted these
densities into Gaussian form and calculated their values
in terms of Gaussian coefficients $c_i$'s and $a_i$'s using relation:
\begin{equation}
\rho(r)=\sum_{i=1}^{n}c_iexp[-a_{i}r^2],
\end{equation}
The Gaussian coefficients which are used as input in the Glauber model
code \cite{mahesh12} are listed in Table 1 for NL3 interaction.

\begin{figure}
%\vspace{-1.7cm}
%\hspace{-0.7cm}
\includegraphics[width=0.8\columnwidth,clip=true]{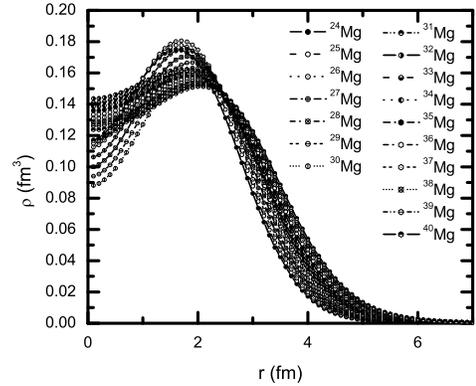}
\label{Fig.1}\vspace{-0.7cm} \caption{The density profile of Mg isotopes from RMF(NL3) formalism as a function of radial distance.}
\end{figure}

\begin{table}
\caption{\label{tab:table3}The Gaussian coefficients for the Projectile and Target
after fitting RMF(NL3) densities  are listed in Table below.}
%\centerline{Table 3 continue.}
\begin{tabular}{cccccc}
\hline \hline
\multicolumn{4}{r}{RMF(NL3)} \\
\cline {2-5}
$Nuclei$& $c_1$& $a_1$ &$c_2$ & $ a_2$\\
\hline
 $^{12}C$ &-0.232654&0.638687&0.517232&0.339911\\
 $^{24}Mg$&-3.6936&0.286163&3.80452&0.261585\\
 $^{25}Mg$&-3.89231&0.281947&4.00019&0.258213\\
 $^{26}Mg$&-4.07533&0.277958&4.18027&0.25493\\
 $^{27}Mg$&-4.065&0.266396&4.1637&0.244635\\
 $^{28}Mg$&-3.93233&0.256062&4.02441&0.234787\\
 $^{29}Mg$&-3.89795&0.246598&3.98355&0.226164\\
 $^{30}Mg$&-3.87287&0.237936&3.95204&0.218274\\
 $^{31}Mg$&-3.41668&0.224842&3.52235&0.206223\\
 $^{32}Mg$&-2.96194&0.211646&3.09244&0.194095\\
 $^{33}Mg$&-2.96276&0.206741&3.09102&0.189597\\
 $^{34}Mg$&-2.96402&0.202127&3.08998&0.185357\\
 $^{35}Mg$&-2.96363&0.197607&3.08718&0.18121\\
 $^{36}Mg$&-2.96752&0.193359&3.08865&0.177322\\
 $^{37}Mg$&-2.96752&0.189208&3.08637&0.173515\\
 $^{38}Mg$&-2.97189&0.185226&3.08797&0.169881\\
 $^{39}Mg$&-2.97547&0.181501&3.08903&0.166466\\
 $^{40}Mg$&-2.98107&0.17785&3.09205&0.163137\\
 \hline \hline
\end{tabular}
\end{table}

For the address of reaction dynamics, the single particle wave function is used in Glauber model.
The radial part of single particle wave function
have been obtained after solving Schrodinger equation using Wood-Saxon type
potential as in the form:
\begin{eqnarray}
U(r)=-v_{0}f(r)+V_{ls}(l.s)r_0^2\frac{1}{r}\frac{df(r)}{dr}+V_{Coul},
\end{eqnarray}
where $f(r)=[1+\exp(\frac{r-R}{a})]^{-1}$ and $R=r_0A^{(1/3)}$. The first term of equation (21) contains
the central potential, second term contains spin orbital part and the last term of the equation contains
Coulomb part of potential. $A$ be the mass number of nucleus. We fixed the value of $r_0=1.2$ fm and
diffuseness parameter $"a"$ as $0.6$ fm in our calculations.\\
Fig. 3 represents the values of $\sigma_{R}$ for $^{24-40}$Mg+$^{12}$C reactions
at $E_{Proj}$=240 AMeV as a function of A of projectile nucleus. The calculated values of
$\sigma_{R}$ in the figure show a remarkable agreement with the experimental
data except for the case of $^{37}$Mg, which support the success of RMF densities for the study
of reaction dynamics. One may see some deviation particularly for higher Mg isotopes and $^{37}$Mg projectile.
This difference in theoretical and experimental data for $^{37}$Mg nucleon is of further significance which seem to
suggest that, it exhibits unusual structure and hence needs further investigation. This point
is explored in further discussion. \\
\begin{figure}
%\vspace{-0.7cm}
%\hspace{-0.7cm}
\includegraphics[width=0.8\columnwidth,clip=true]{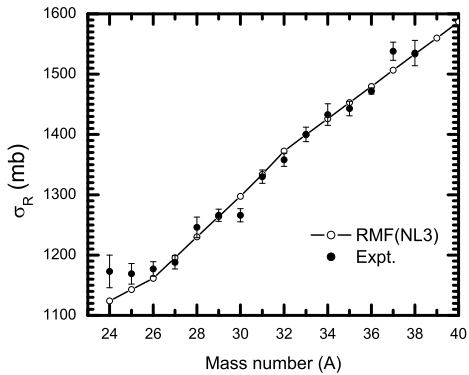}
\label{Fig.1}\vspace{-0.7cm} \caption{ Reaction cross section for $^{24-40}$Mg as projectile
with $^{12}$C target nucleus at $E_{proj}$= 240 AMeV. The experimental data are also given for comparison \cite{takechi2015}.}
\end{figure}

Fig. 4 shows the angular elastic differential cross sections of $^{34-38}$Mg
projectiles on the carbon target at $E_{Proj}$=240 AMeV. The inspection of the figure suggest
that, two dip positions are observed at an angles $\theta_{c.m}$ = $2^0$ and $8^0$. One may also observe that the
large dip appeared for $^{37}$Mg and $^{38}$Mg at these angles. Specifically speaking, the
largest dip at $^{37}$Mg projectile may be associated with the loosely bound structure of this nucleus.
The above observations give indication of halo behavior of $^{37}$Mg.\\

\begin{figure}
\vspace{-0.7cm}
\hspace{-0.7cm}
\includegraphics[width=0.8\columnwidth,clip=true]{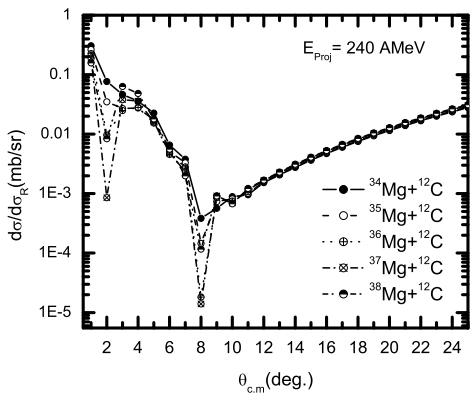}
\label{Fig.1}\vspace{-0.7cm} \caption{The values of angular elastic differential cross
section for $^{34-38}$Mg+$^{12}$C reactions at $E_{proj}$= 240 AMeV.}
\end{figure}

\begin{figure}
%\vspace{-0.7cm}
%\hspace{-0.7cm}
\includegraphics[width=0.8\columnwidth,clip=true]{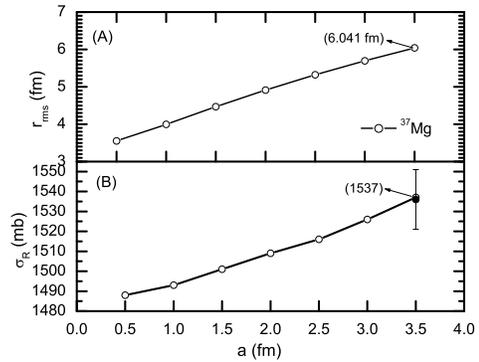}
\label{Fig.1}\vspace{-0.7cm} \caption{(A) Upper panel of the figure show the rms
radius ($r_{rms}$) in fm and (B) lower panel show the values of reaction cross section
$\sigma_R$ in mb as a function diffuseness parameter 'a' in fm for $^{37}$Mg.}
\end{figure}

For further study of the case of $^{37}$Mg projectile,
we have used Glauber formalism with two body (core+nucleon) system.
Further details of the calculations can be seen in \cite{Abu2003,mahesh13}.
Fig. 5 represents the calculated values of $\sigma_R$ (in mb) and $r_{rms}$
(in fm) for the
$^{37}$Mg projectile. The upper panel of Fig. 5 presents the root mean square radius $r_{rms}$
and lower panel represents $\sigma_{R}$ for $^{37}$Mg as a projectile
over the $^{12}$C target at energy 240 AMeV as a function of diffuseness parameter 'a' in fm.
We have tried to fit the reaction cross section for $^{37}$Mg projectile at different values of
diffuseness parameter. It is clear from the figure that the reaction cross section and diffuseness parameter
are linearly dependant to each other. The value of $\sigma_R$ obtained at diffuseness parameter = 3.5 fm is 1537 mb,
which is well comparable to the experimental observation 1536$\pm$15 mb. Hence the lower panel of the
figure shows that the value of reaction cross
section fit with the experimental value at a = 3.5 fm. The upper panel of the figure shows the rms radius
values of core+neutron system as a function of diffuseness parameter. We observed that the value of root mean square radius of
projectile (core + nucleon) at a=3.5 fm is 6.041 fm. Thus the large value of reaction
cross section has direct consequence to their large radius and halo nature of $^{37}$Mg isotopes.

\begin{figure}
%\vspace{-0.7cm}
%\hspace{-0.7cm}
\includegraphics[width=0.8\columnwidth,clip=true]{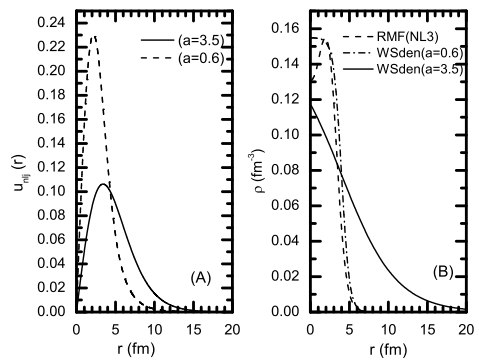}
\label{Fig.1}\vspace{-0.7cm} \caption{(A) Comparison of single particle wave function for different values of diffuseness parameter 'a' and (B) Comparison of the Wood Saxon densities with different diffusion parameter and RMF(NL3) density of $^{37}$Mg as function of radial distance in fm.}
\end{figure}
\begin{figure}
%\vspace{-0.7cm}
%\hspace{-0.7cm}
\includegraphics[width=0.8\columnwidth,clip=true]{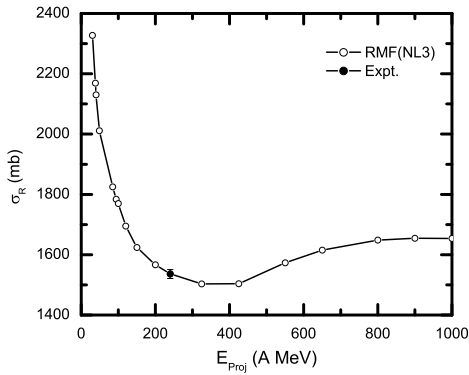}
\label{Fig.1}\vspace{-0.7cm} \caption{Variation of Total reaction cross sections for $^{37}$Mg as function of $E_{Proj}$. }
\end{figure}
\begin{figure}
%\vspace{-0.7cm}
%\hspace{-0.7cm}
\includegraphics[width=0.8\columnwidth,clip=true]{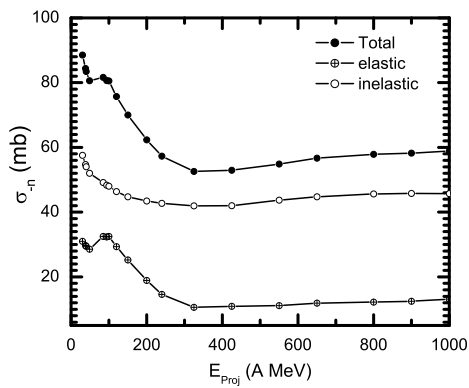}
\label{Fig.1}\vspace{-0.7cm} \caption{One neutron removal cross section for the reaction $^{37}$Mg+$^{12}$C reaction including elastic and inelastic part as a function of $E_{Proj}$ A MeV}
\end{figure}
\begin{figure}
%\vspace{-0.7cm}
%\hspace{-0.7cm}
\includegraphics[width=0.8\columnwidth,clip=true]{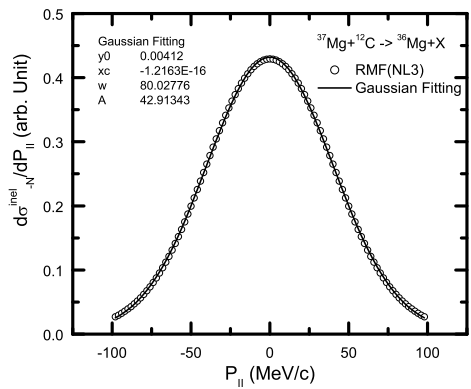}
\label{Fig.1}\vspace{-0.7cm} \caption{Longitudinal momentum distribution of $^{36}$Mg from the $^{37}$Mg+$^{12}$C at projectile energy 240 A MeV.}
\end{figure}
Fig.6 represents the comparisons of wave function for different value of diffuseness parameters and
RMF density for $^{37}$Mg with Wood-Saxon density having diffuseness parameter 0.6 fm and 3.5 fm.
The left side of the figure show variation of single particle wave function with radial distance for different values of diffuseness parameter.
This figure suggest that the wave function is more steeper at diffuseness value of 0.6 fm where as more broader at 3.5 fm.
The right side of this figure show that the behavior of Wood-Saxon density is similar to RMF at a=0.6 fm.
At the value of 'a = 3.5 fm', we get a long extension of density even beyond $\sim$ 15 fm, further suggesting the possibility of
a halo structure of $^{37}$Mg nucleus. Although, this large value of a = 3.5 fm
looks unusual, but still be relevant for exotic nuclei exhibiting halo structure.
For further significance of higher diffuseness parameter "a", one need to have extensive investigation
by addressing reaction dynamics involving various halo nuclei. We have fitted density with a
Wood-Saxon form taking variation in diffuseness
parameter 'a'. Then, we used these densities in the reaction calculation to evaluate the reaction cross-section $\sigma_R$. We find that the fitted
density reproduce the experimental data for $^{37}$Mg at a = 3.5 fm.\\
Fig. 7 represents the variation of reaction
cross section as a function of projectile energy for 30-1000 AMeV using wave function with a=3.5. The experimental values are also
given for comparison. Fig. 8 shows the one neutron removal cross section for $^{37}$Mg+$^{12}$C
reaction as a function of $E_{Proj}$. One neutron removal cross section consists of both elastic
and inelastic component. It is clear from the figure that the inelastic component in neutron removal
cross section dominate over their elastic component at $E_{Proj}$=240 A MeV. The trend of reaction cross
section in figure 7 and one neutron removal cross section in figure 8 are similar, but small hike is appeared
in one neutron removal cross section at $E_{Proj}$=100 A MeV, because of its elastic component.\\
Fig. 9 show the calculated longitudinal momentum distribution of $^{36}$Mg core for the reaction $^{37}$Mg+$^{12}$C
at $E_{Proj}$=240 A MeV. The trend of distribution exhibits the Gaussian pattern. So we compare our calculated values
of longitudinal momentum distribution using RMF densities with in Gaussian function.
%\begin{equation}
%Y=Y_0+\frac{A}{w.\sqrt{\frac{\pi}{2}}}e^{-\frac{2(x-x_0)^2}{w^2}}.
%\end{equation}
%%Where $Y_0$ is the base line $X_0$ is the center of peak and 'A' is the Area under the base
%line and 'W' is the width of FMHW.
The circle points show the calculated values of momentum distribution of one nucleon from the $^{37}$Mg
projectile and solid line is the fitted gaussian curve. By fitting curve the observed value of FWHM comes out to be 80.02
MeV/c.
\section{SUMMARY}
In summary, we have calculated the ground state properties of Mg isotopes and
also studied the reaction cross sections of these isotopes taken as projectile from
the valley of stability to drip line region with stable $^{12}$C target at $E_{proj}$ 240 AMeV.
We found remarkable agreement of ground state properties of Mg-isotopes with available data. The estimated values of
reaction cross section using densities from RMF formalism are nicely compered with the experimental data. The excellent
agreement of estimated reaction cross section values except for $^{37}$Mg isotope is an evidence of predictive power
of RMF.  The skin effect variation with mass number is
studied in context of density profile. The study of angular elastic differential cross section
for $^{34-38}$Mg and further investigation with Glauber two body calculation also support the halo status of $^{37}$Mg.
Subsequently we examined the halo status of $^{37}$Mg and it seems justified from its higher
magnitude of rms radius $\sim$6.041 fm and small value of FWHM (80.02 MeV/c) of
longitudinal momentum distribution.
%\section*{Acknowledgments}

\end{document}